\documentclass[fleqn,10pt]{SelfArx} 

\definecolor{color1}{RGB}{0,0,90} 
\definecolor{color2}{RGB}{0,20,20} 

\usepackage{hyperref} 
\hypersetup{hidelinks,colorlinks,breaklinks=true,urlcolor=color2,citecolor=color1,linkcolor=color1,bookmarksopen=false,pdftitle={Title},pdfauthor={Author},urlcolor=blue}

\usepackage{graphicx}
\usepackage[square]{natbib}
\usepackage{amsmath}
\usepackage{multirow}
\usepackage{subfigure}
\usepackage{pdflscape}

\newcommand{\n}[1]{\mathrm{#1}}

\JournalInfo{Published in Journal of Magnetism and Magnetic Materials, Vol. 465, 106-113, 2018} 
\Archive{\href{https://doi.org/10.1016/j.jmmm.2018.05.076}{DOI: 10.1016/j.jmmm.2018.05.076}} 

\PaperTitle{A topology optimized switchable permanent magnet system} 

\Authors{R. Bj\o{}rk$^1$ and A. R. Insinga$^1$} 
\affiliation{\textsuperscript{1}\textit{Department of Energy Conversion and Storage, Technical University of Denmark - DTU, Frederiksborgvej 399, DK-4000 Roskilde, Denmark}} 
\affiliation{*\textbf{Corresponding author}: rabj@dtu.dk} 

\newcommand{\keywordname}{Keywords} 

\Abstract{The design of a magnetic field source that can switch from a high field to a low field configuration by rotation by $90^\circ$ of a set of iron pieces is investigated using topology optimization. A Halbach cylinder is considered as the magnetic field source and iron inserts are placed in the air gap of the Halbach cylinder. The ideal shape of these iron inserts is determined as function of the field generated by the Halbach cylinder and as function of the size of the iron segments. The topology optimized structures are parabolic shaped pieces and have a difference in flux density between the high and low positions that is on average 1.29 times higher than optimized regular pole pieces. The maximum increase is a factor of 2.08 times higher than the regular pole pieces.}

\begin{document}

\flushbottom 

\maketitle 


\thispagestyle{empty} 

\section{Introduction}
A permanent magnetic field source that can easily be switched ``on and off'' would be of significant interest for e.g. nuclear magnetic resonance (NMR) apparatus \cite{Appelt_2006} and magnetic refrigeration devices \cite{Smith_2012}. Such variable permanent flux sources have previously been investigated \cite{Abele_1993,Coey_2002,Bjoerk_2010g} and several designs have been proposed that can alternate between a high and low field \cite{Cugat_1994,Mhiochain_1999,Hilton_2012}. However, these designs all involve moving permanent magnets or contain a large number of complex-shaped permanent magnets that are not easily realizable.

Here we wish to investigate if a standard permanent magnetic field source such as a Halbach cylinder can be turned into a variable field source, that can be switched from a high to a low field, by modifying the structure with inserts of iron. The magnetic field is created in a circular bore and the inserts are placed in this bore, and are rotated to change the field. This can in some regard already be done using pole pieces that are rotated depending on the field desired. In order to investigate if a more efficient design of inserts exist, topology optimization is here employed.

Topology optimization is traditionally used for structural mechanics \cite{Bendsoe_2013}, but have also recently been used to design permanent magnet systems. Specifically, topology optimization has been used to determine the optimal direction of the magnetization \cite{Wang_2005,Choi_2012}, for designing C-core actuators \cite{Choi_2012,Lee_2012} and to optimize permanent magnet structures in general \cite{Bjoerk_2017,Lee_2018}. Finally, topology optimization has also been used extensively in optimization of electrical motors \cite{Wang_2005,Choi_2014,Ishikawa_2015,Putek_2016} and pole pieces for MRI systems \cite{Lee_2010,Tadic_2011}.

\begin{figure*}[!t]
  \centering
  \includegraphics[width=2\columnwidth]{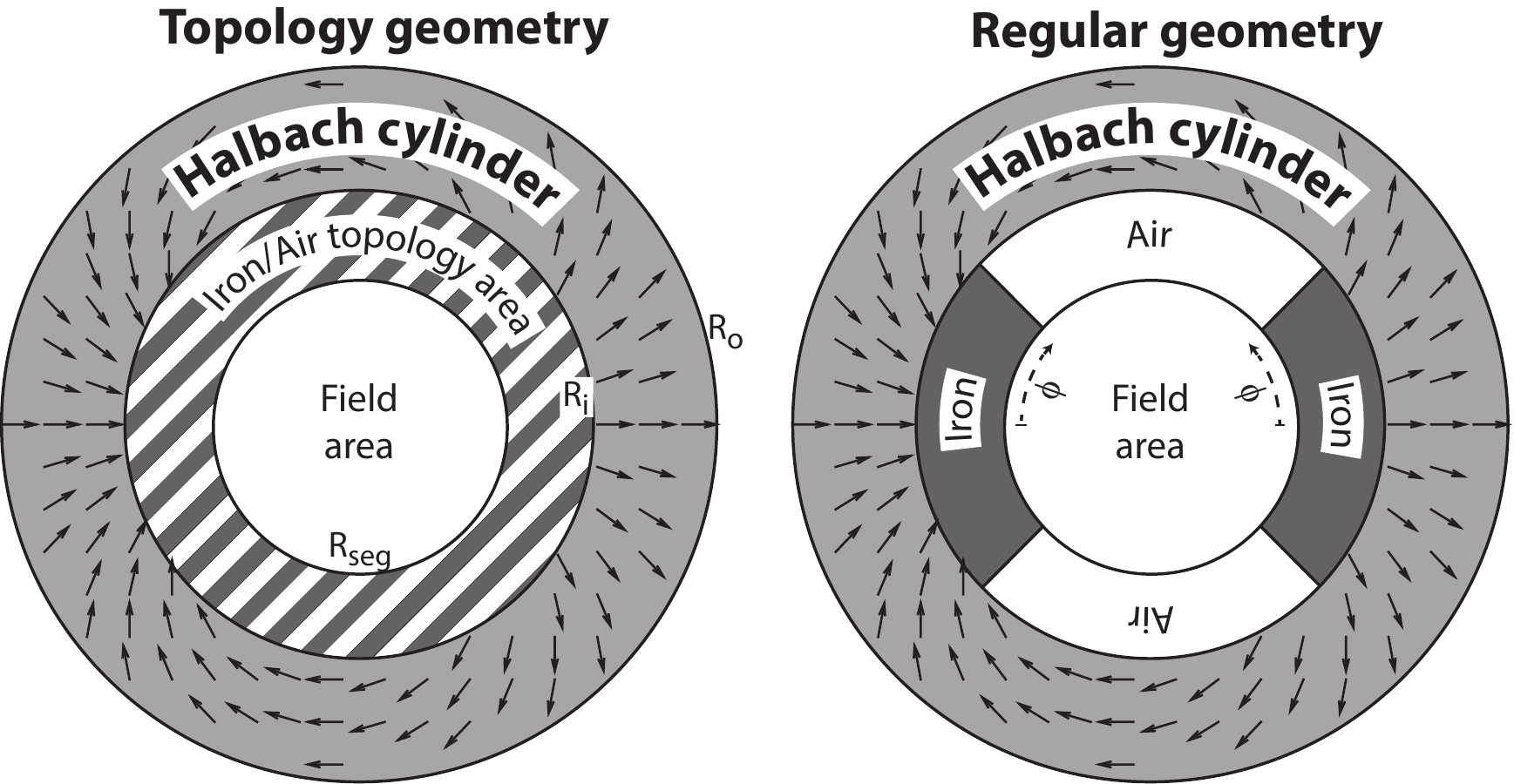}
  \caption{An illustration of the geometry considered. The left geometry is the Halbach cylinder with the structure to be topology optimized, while the right hand structure is the regular geometry used for comparison. Note the angle $\phi$, which is the angular extent of the iron inserts for the regular geometry. The iron inserts are rotated 90$^\circ$ to change the field from high to low field in all cases.}
  \label{Fig_Switch_Halbach_ill}
\end{figure*}

Here we will consider a problem where the optimal shape of iron inserts must be determined, in order to provide the greatest difference in field when the iron inserts are rotated. The structure of permanent magnets are considered fixed, in the form of a Halbach cylinder. The present work constitutes the first topology optimization study that considers an objective which must necessarily be formulated in terms of the field distribution corresponding to two different states. The geometry to be topology optimized is compared with a regular pole piece geometry consisting of two arc-shaped iron segments. For this geometry the angular span of the iron segments are varied until the largest difference in average field is produced. The geometry of this system and of the system to be topology optimized is shown in Fig. \ref{Fig_Switch_Halbach_ill}.

\section{Method and implementation}
The numerical model used for calculating the magnetic field as well as to perform the topology optimization is the finite element framework Comsol Multiphysics. Specifically, the Globally Convergent Method of Moving Asymptotes (GCMMA) solver is used \cite{Svanberg_1987,Svanberg_2002}, as this is well suited to topology optimization due to its ability to solve problems with a large number of control variables. This topology implementation is the standard implementation used in Comsol Multiphysics. In general the computational framework used is the same as in Bj\o{}rk et al. (2017) \cite{Bjoerk_2017}, and the same number of mesh elements has been used in this work. In general in topology optimization the distribution of materials throughout parts of the geometry of the problem is considered as the design variable.

The global objective to be optimized for, termed $\Theta$, using topology optimization must depend on the topology of the system. We wish to determine the shape (topology) of the iron segments that, when rotated $90^\circ$ inside the Halbach cylinder, results in the highest change in magnitude of the field in the cylinder bore. In other words the field should ideally change from a large value to a small value, when the iron segments are rotated. When the iron segments are parallel to the magnetic field generated in the cylinder bore, the field lines are focused in the central region, where the intensity of the field increases. On the other hand when the iron inserts are placed perpendicular to the field in the bore, the field lines may bypass the central region, thus reducing the field intensity inside it. The objective of the topology optimization, $\Theta$, is the difference in the average field with the iron segments rotated $0^\circ$ and $90^\circ$ respectively, i.e.
\begin{eqnarray}
\Theta = \langle{}B\rangle{}_{0^\circ}-\langle{}B\rangle{}_{90^\circ} = \langle{}B\rangle{}_{high}-\langle{}B\rangle{}_{low}
\end{eqnarray}

We consider topology optimization between two materials: high permeability iron and air. The high permeability iron has a non-linear $B-H$ curve as provided in the Comsol material library. This material data is shown in Fig. \ref{Fig_Iron_curve_mur} in the Appendix. The saturation magnetization is around 2 T. Previous studies on non-linear materials have shown that it is important to account for the full $B-H$ curve of the materials \cite{Lee_2012}. The permanent magnet material, which is considered fixed, has a linear $B-H$ relation with a permeability of $\mu_r=1.05$ and a remanence of 1.4 T. However, as long as the permeability is close to 1 the resulting magnetic field in the bore is proportional to the remanence, which means that the system can be parameterized by the Halbach formula for the field in the bore $B=B_\n{rem}\n{ln}(R_\n{o}/R_\n{i})$.

In the topology optimization problem a design region must be split into regions of two different distinct magnetic materials. In order to obtain a sharp geometrical transition between the different material types, a penalty function of a single topology control variable, $q$, is introduced that can switch between two material types. The control variable has a range between 0 and 1 and is a function of position. In order to change between iron and air, the relative permeability is changed. This is locally given as
\begin{equation} \label{Eq.Mur_penalty}
\mu_r = \left(1-\frac{1+\mu_{r,\n{iron}}}{2}\right)\mathrm{tanh}\left(\frac{q-0.5}{0.01}\right)+\frac{1+\mu_{r,\n{iron}}}{2}
\end{equation}
where $\mu_{r,\n{iron}}$ is the relative permeability of iron. Note that $\mu_{r,\n{iron}}$ is a function of the norm of the magnetic field, $H$. Here if $q=0$ the material has the relative permeability equal to $\mu_{r,\n{iron}}$, while if $q=1$, the relative permeability is $\mu_{r} = 1$. A somewhat similar approach was used in Refs. \cite{Bjoerk_2017,Choi_2014,Lee_2010}.

Specifically, only a quarter of the geometry shown in Fig. \ref{Fig_Switch_Halbach_ill} is modelled. The appropriate boundary conditions (making the field parallel or orthogonal to the boundary) are chosen based on the orientation of the iron pieces, i.e. whether they are in the high or low field position. The outer boundary of the simulation, which is located far from the outer boundary of the Halbach cylinder is magnetically insulating.

\section{Field source}
The permanent magnetic field source is here taken to be an ideal Halbach cylinder, as this provides a completely homogeneous magnetic field in the cylinder bore. A Halbach cylinder is used as this is a structure that is relatively easy to produce and is already used in a large number of applications such as nuclear magnetic resonance (NMR) equipment \cite{Appelt_2006,Moresi_2003}, magnetic refrigeration devices \cite{Tura_2007,Bjoerk_2010b} and medical applications \cite{Sarwar_2012}. Using topology optimization it is also possible to optimize for the direction of the remanence of the permanent magnet field source \cite{Choi_2012,Choi_2014,Lee_2018}, which may lead to a higher performing device, but typically at the cost of a greater complexity. Here we have chosen to use a fixed remanence, that can be relatively easily manufactured. The polar components of the remanence of a Halbach cylinder are given as \cite{Halbach_1980}
\begin{eqnarray}\label{Eq.Halbach}
B_\n{rem,r} &=& B_\n{rem}\cos(p\phi)\hat{r} \nonumber\\
B_\n{rem,\phi} &=& B_\n{rem}\sin(p\phi)\hat{\phi}
\end{eqnarray}
Here $B_\n{rem}$ is the remanence, and we take $p=1$, so that the field created in the cylinder bore is perfectly homogeneous as long as the bore of the magnet is completely filled with air, or any other non-magnetic material. For a Halbach cylinder with $p=1$ the magnitude of the field in the bore will be given by $B=B_\n{rem}\mathrm{ln}\left(R_\n{o}/R_\n{i}\right)$, where $R_\n{o}$ is the outer radius and $R_\n{i}$ is the inner radius of the cylinder \cite{Halbach_1980}. This expression is here an approximation, as the permeability of the magnets are not 1, but the previously mentioned 1.05. However, this is sufficiently close to 1, that the expression for the field magnitude can be used. For a larger value of $\mu_r$, equation 31 from Bj\o{}rk et al. 2010 applies \cite{Bjoerk_2010a}. We assume an infinitely long cylinder, and that there is no self-demagnetization in the Halbach cylinder \cite{Bjoerk_2015a,Insinga_2016a}. In other words, since the behavior of the permanent magnet is modeled using a linear $B$-$H$ relation, the field generated by the Halbach cylinder is always perfectly homogeneous, and its magnitude is given by the expression reported above. The validity of this assumption is discussed in Section \ref{sec:Coercivity} and a segmented Halbach cylinder is considered in Section \ref{sec:Segmented}.

It is advantageous to consider the results in terms of the logarithmic equation given in the above paragraph because this means that the results to be applied to a Halbach cylinder with an arbitrary remanence and radii. In other words, the geometry of the iron insert will be the same for a Halbach cylinder with a remanence of 1.0 T and a ratio of the radii of 2.5 and a Halbach cylinder with a remanence of 1.4 T and a ratio of the radii of 1.924, because these produce the same field in the bore. Note however that placing the inserts in the bore will increase the field. This is discussed subsequently.

\section{Results}
The optimal shape of the iron segments inside the cylinder bore is calculated as function of two parameters, namely the inner radius of the iron segments, and the outer radius of the Halbach cylinder, which determines the field strength, for the given value of the remanence. The first parameter controls the available area of iron to guide the flux lines, while the second parameter controls the flux density generated in the cylinder bore, i.e. the field that must be guided. The ratio of the inner radius of the iron segment, $R_\n{seg}$, to the inner radius of the Halbach cylinder, $R_\n{i}$, was varied from 0.2 to 0.8 in steps of 0.1, while the ratio of the outer and inner radius of the Halbach cylinder, $R_\n{o}/R_\n{i}$, was varied from 1.2 to 3.0 in steps of 0.2. The models are first solved on a somewhat coarse mesh, which is then refined twice in subsequent simulations. The typical computation time including all mesh refinements is $\sim{}24$ hours on an AMD Ryzen Threadripper 1950X.

\begin{figure}[!t]
  \centering
  \includegraphics[width=1\columnwidth]{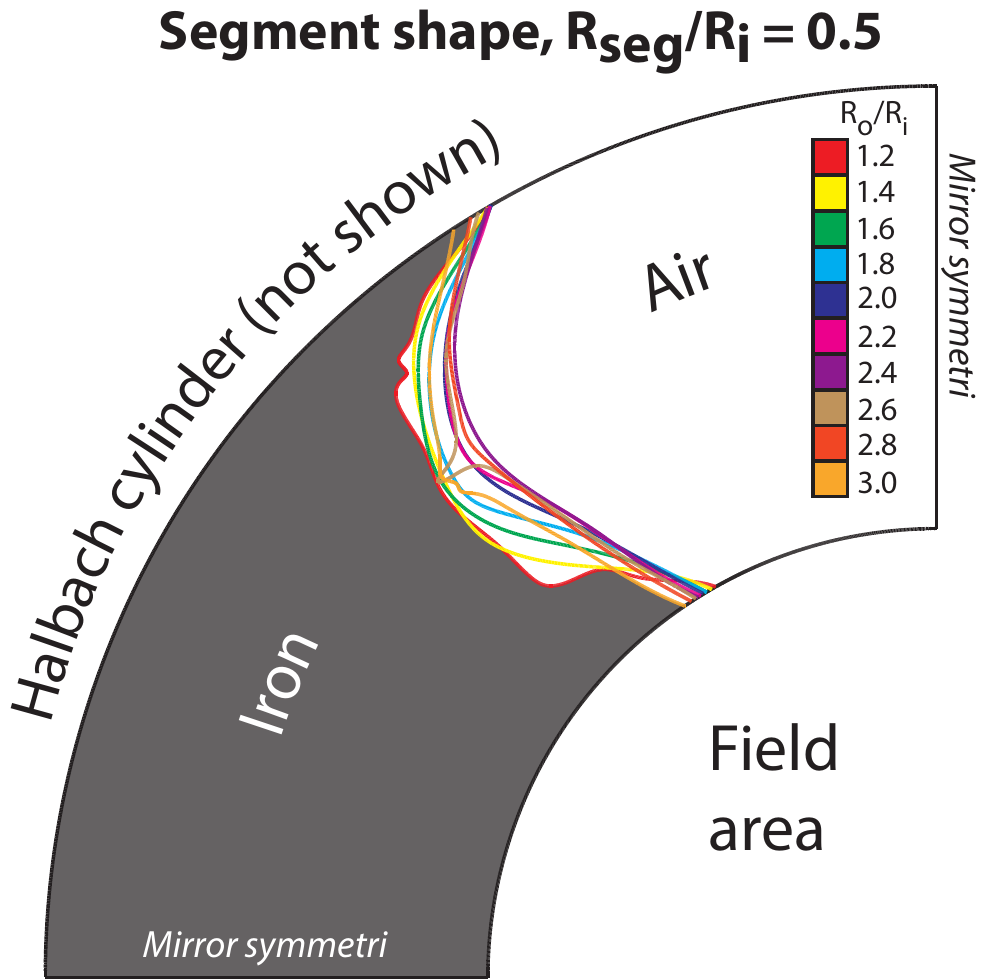}
  \caption{The shape of a quarter of the topology optimized segment as function of $R_\n{o}/R_\n{i}$ for $R_\n{seg}/R_\n{i}=0.6$. There is mirror symmetry along the axes.}
  \label{Fig_Illus_switch_insert_RoRi}
\end{figure}

\begin{figure*}[!t]
  \centering
  \includegraphics[width=2\columnwidth]{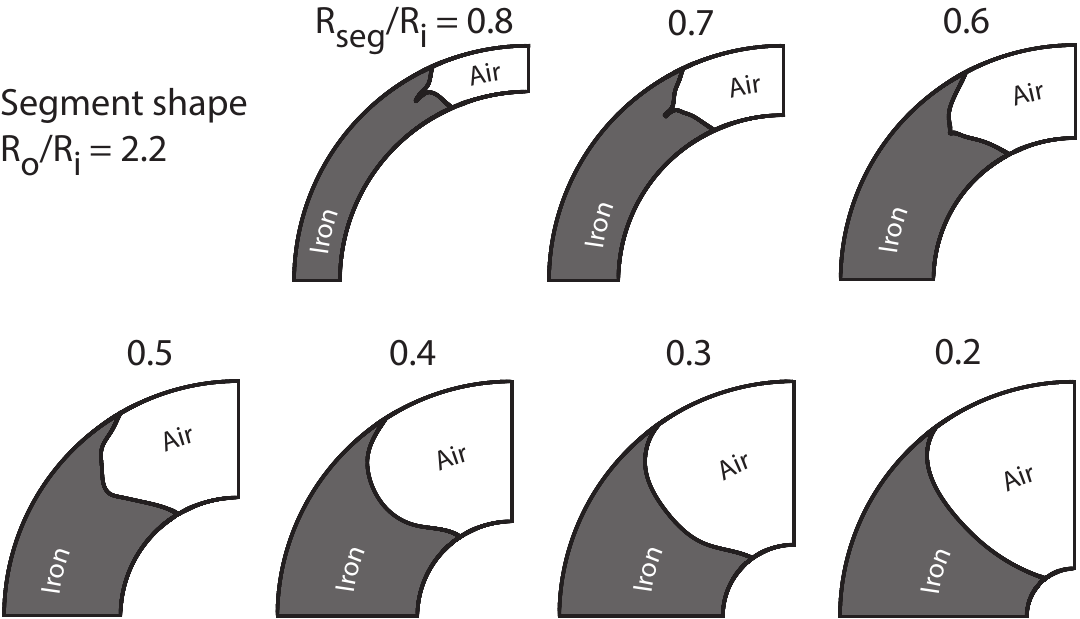}
  \caption{The shape of a quarter of the topology optimized segment as function of $R_\n{seg}/R_\n{i}$ for $R_\n{o}/R_\n{i}=0.2$. There is mirror symmetry along the axes.}
  \label{Fig_Illus_switch_insert_RsegRi}
\end{figure*}

The result of the topology optimization is a map of the optimal permeability, specified through the topology variable $q$ in Eq. \ref{Eq.Mur_penalty}. This map of the permeability must be turned into a physical object, namely the insert, i.e. the boundaries of the insert must be defined. We take the border of the iron and air region as the contour curve of the topology optimized geometry for which $\mu_r=1.1$. The shape of these iron segments are shown in Fig. \ref{Fig_Illus_switch_insert_RoRi} as function of $R_\n{o}/R_\n{i}$ for $R_\n{seg}/R_\n{i}=0.6$, while they are shown as function of $R_\n{seg}/R_\n{i}$ for $R_\n{o}/R_\n{i}=0.2$ in Fig. \ref{Fig_Illus_switch_insert_RsegRi}. As can be seen from the figures, the topology that maximises the difference in average field have a somewhat parabolic surface between the iron and the bordering air.

\begin{figure*}[!t]
  \centering
  \includegraphics[width=2\columnwidth]{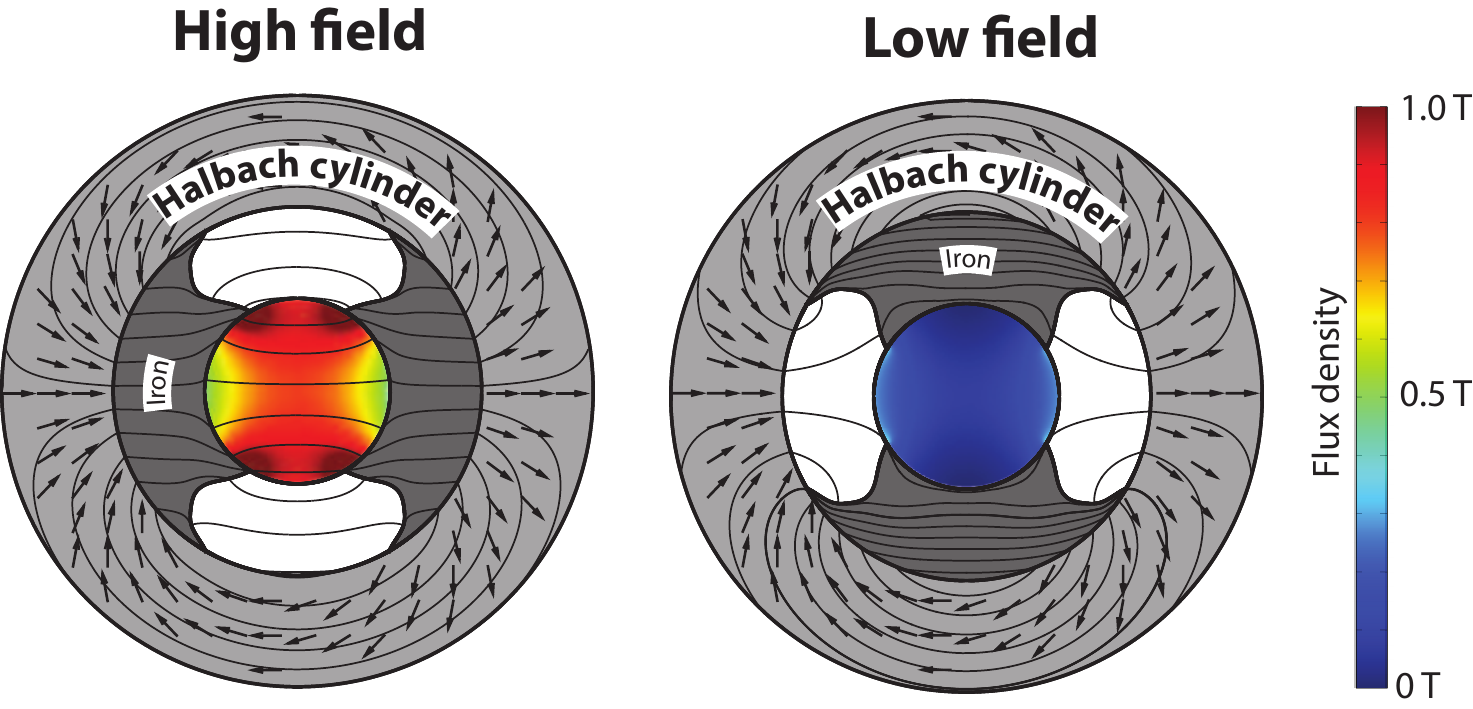}
  \caption{The high and low field configuration for a $R_\n{o}/R_\n{i}=1.6$ and $R_\n{seg}/R_\n{i}=0.5$ structure with $B_\n{rem}=1.4$ T.}
  \label{Fig_Full_14_0.05_16_ill.eps}
\end{figure*}

An illustration of the flux density in the high and low field configurations are shown in Fig. \ref{Fig_Full_14_0.05_16_ill.eps} for a configuration with $R_\n{o}/R_\n{i}=1.6$ and $R_\n{seg}/R_\n{i}=0.5$.

\begin{figure}[!t]
  \centering
  \includegraphics[width=1\columnwidth]{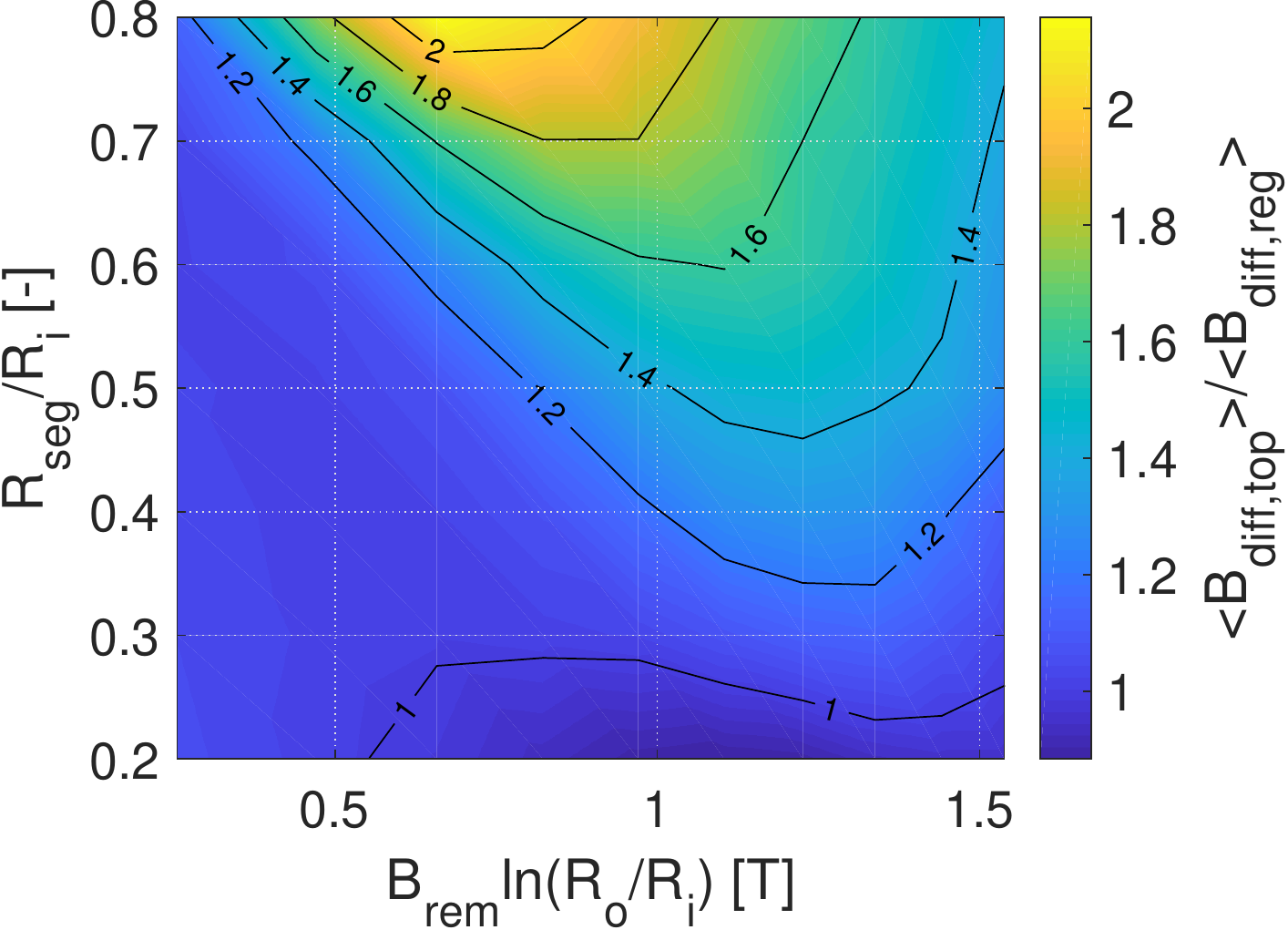}
  \caption{The ratio of the difference in average high and low field of the topology optimized geometry, $<B_\n{diff,top}>$, and the regular geometry, $<B_\n{diff,reg}>$, as function of the relative radius of the magnet and the switch.}
  \label{Fig_B_diff_con_cut}
\end{figure}

The ratio of the difference in average field for the topology optimized geometry and the regular geometry is shown in Fig. \ref{Fig_B_diff_con_cut}. The results are shown as function of the ratio between $R_\n{seg}$ and $R_\n{i}$ as well as the field in the central bore, which is given as $B=B_\n{rem}\n{ln}(R_\n{o}/R_\n{i})$, thus here the remanence is accounted for. We have run a complete set of topology optimization with a remanence of 1 T and 1.4 T and verified that the results are identical, once the results are shown as function of internal field in the Halbach, as above. Note that for the regular geometry, the angular extend of the segments, $\phi$, have also been optimized for every set of parameters. The optimal angle as function of  the ratio between $R_\n{seg}$ and $R_\n{i}$ and the field in the central bore is shown in the Appendix in Fig. \ref{Fig_Angle_14_cut}.

\begin{figure}[h]
\begin{center}
\subfigure[]{}\includegraphics[width=0.47\textwidth]{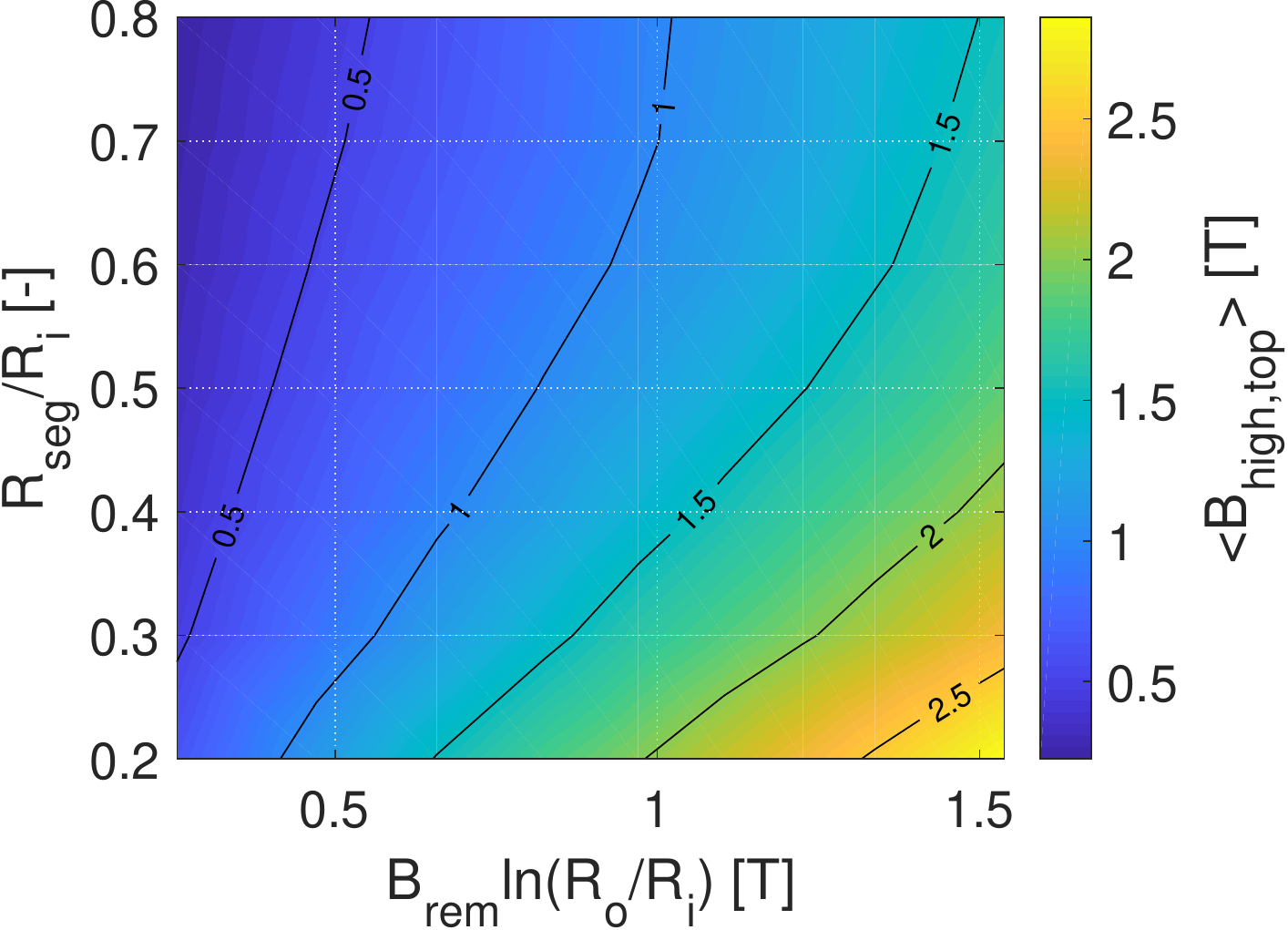}
\subfigure[]{}\includegraphics[width=0.47\textwidth]{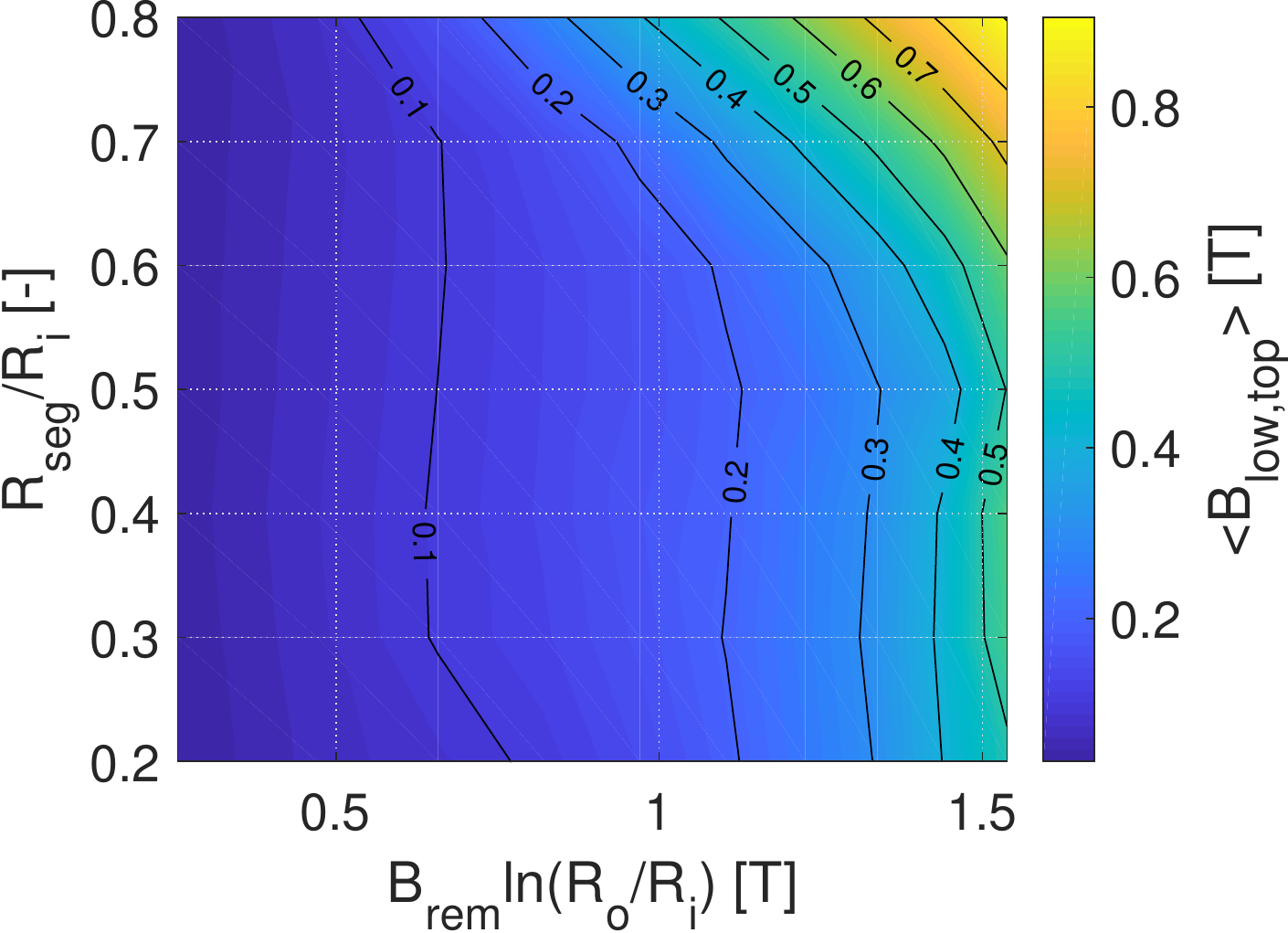}
\end{center}
\caption{a) The average high field and b) the average low field for the topology optimized structure as function of the relative radius of the magnet and the switch.}\label{Fig_B_top}
\end{figure}

The topology optimized structures have a difference in flux density between the high and low configurations that is on average 1.29 times higher than the optimized regular pole pieces. The maximum increase is a factor of 2.08 times higher than the regular pole pieces. The average high field (corresponding to a rotation of the segments of $0^\circ$) and the average low field (corresponding to a rotation of $90^\circ$) are shown in Fig. \ref{Fig_B_top}.
Note that when the iron inserts are placed in the bore, the field is increased significantly. This can be seen from Fig. \ref{Fig_B_top}a, which shows the average magnetic field in the high field position. As can be seen, the field generated is significantly higher that the base field of the Halbach, which is given on the $x$-axis of Fig. \ref{Fig_B_top}a.

\subsection{Parameterizations of the segment shape}
The shape of the border of the topology optimized insert is similar to a parabola. Therefore, second order polynomials was fitted to the shape of the insert for all geometry parameters. The inserts were normalized to extend from a radius of zero to one, when fitting, which in non-normalized coordinates corresponds to $R_\n{seg}$ and $R_\n{i}$, respectively. We assume that the second-order polynomial is mirror symmetric with respect to the middle of the insert. This means that the first coefficient of the second order polynomial is equal to minus the second coefficient. Thus the second order polynomials fitted in polar coordinates are
\begin{eqnarray}
\phi = -k_1r^2+k_1r+k_0
\end{eqnarray}
where $\phi$ is the angular coordinate of the insert expressed in radians as function of the normalized radial coordinate $r$.

\begin{figure*}[!t]
\begin{center}
\subfigure[]{}\includegraphics[width=0.47\textwidth]{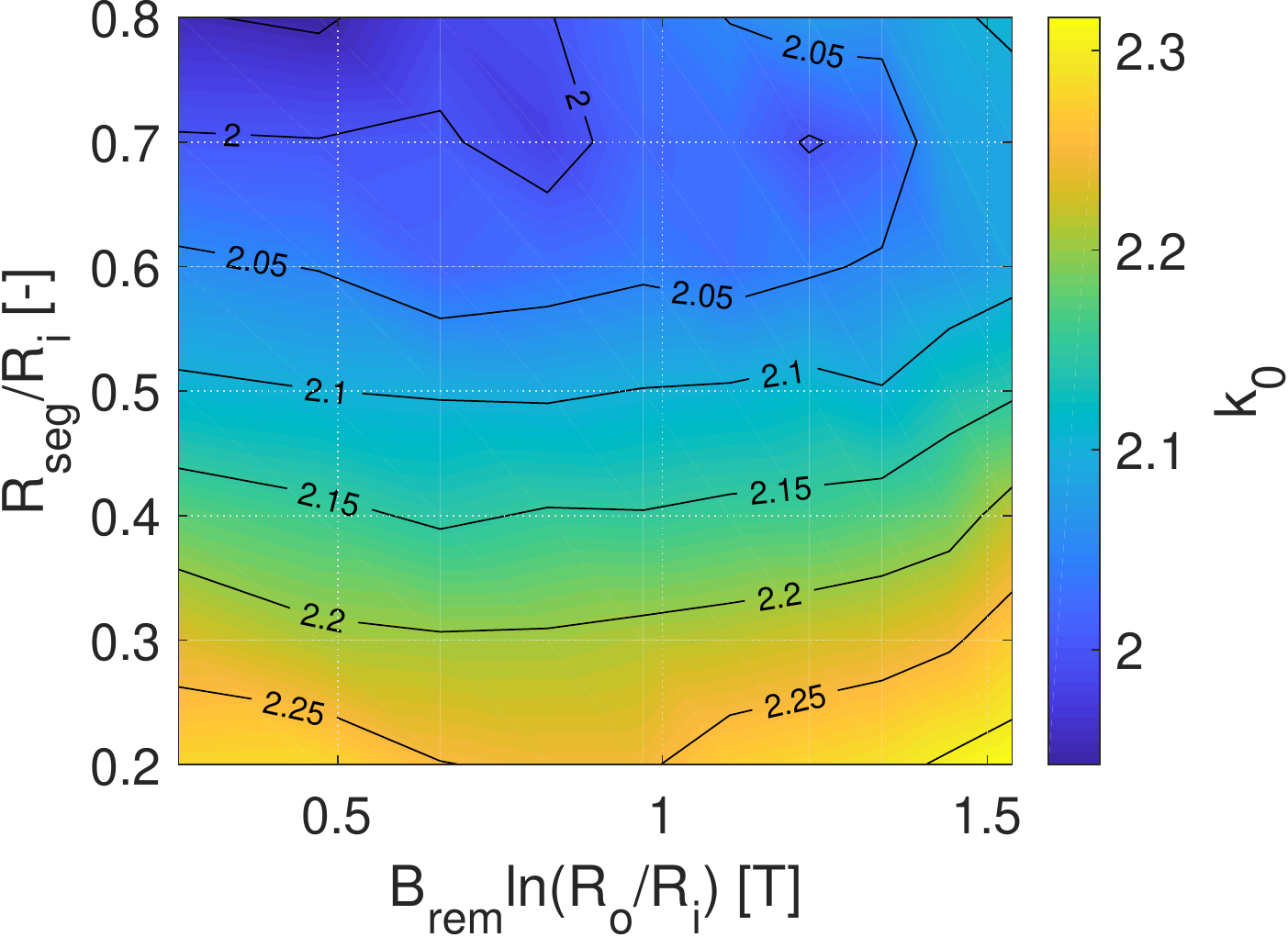}
\subfigure[]{}\includegraphics[width=0.47\textwidth]{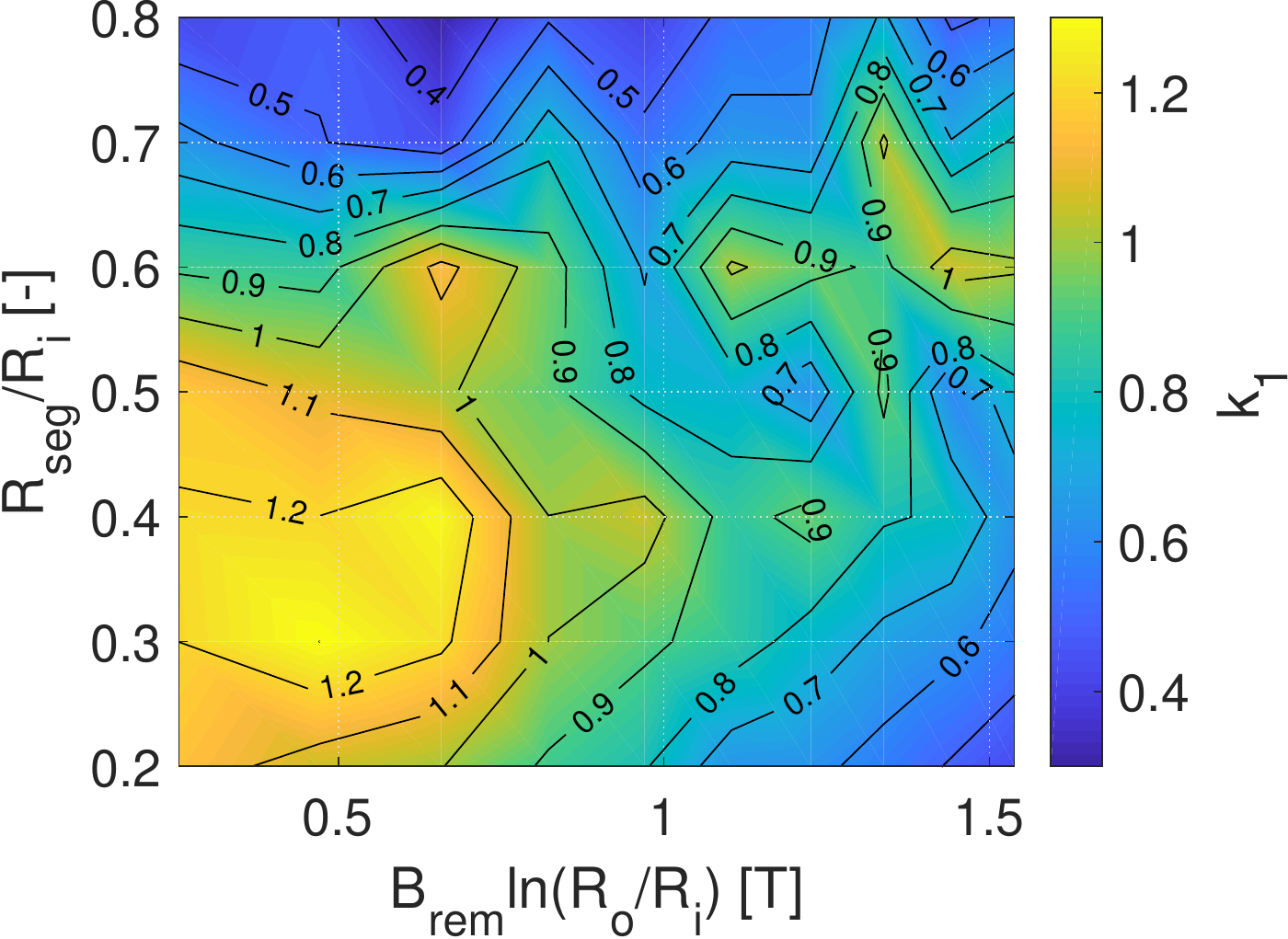}
\end{center}
\caption{The coefficients of the parabolic fit to the shape of the segment as function of the relative radius of the magnet and the switch.}\label{Fig_p_coeff}
\end{figure*}

The fitted values of the polynomial coefficients with geometry is shown in Fig. \ref{Fig_p_coeff}. For these parabolic shaped inserts, the difference in flux density between the high and low field configurations is on average 1.29 times higher that the optimized regular pole pieces. This ratio is the same as for the topology optimized structures, i.e. there is no difference between the parabolic and the topology optimized structures.

\begin{figure*}[!t]
\begin{center}
\subfigure[]{}\includegraphics[width=0.47\textwidth]{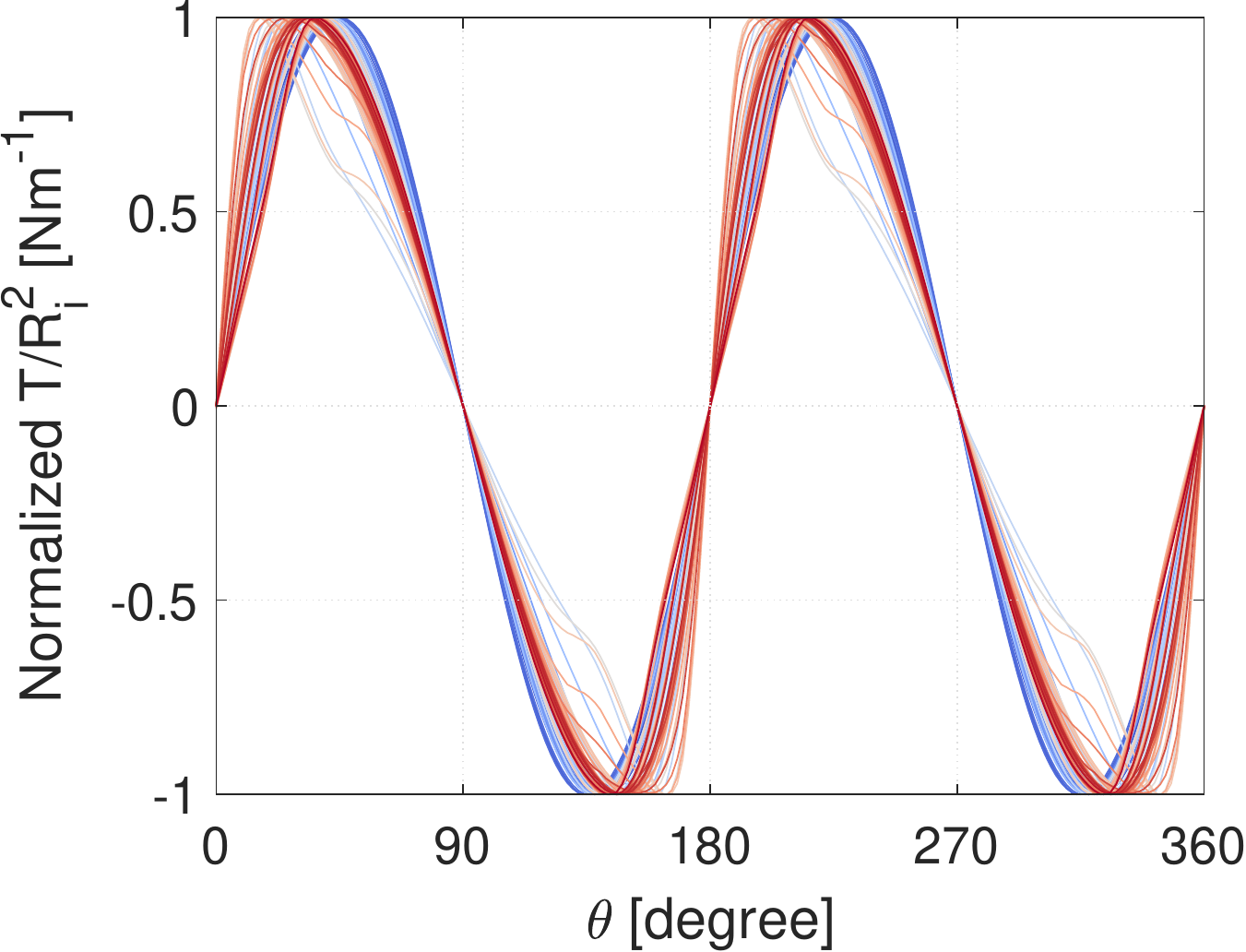}
\subfigure[]{}\includegraphics[width=0.47\textwidth]{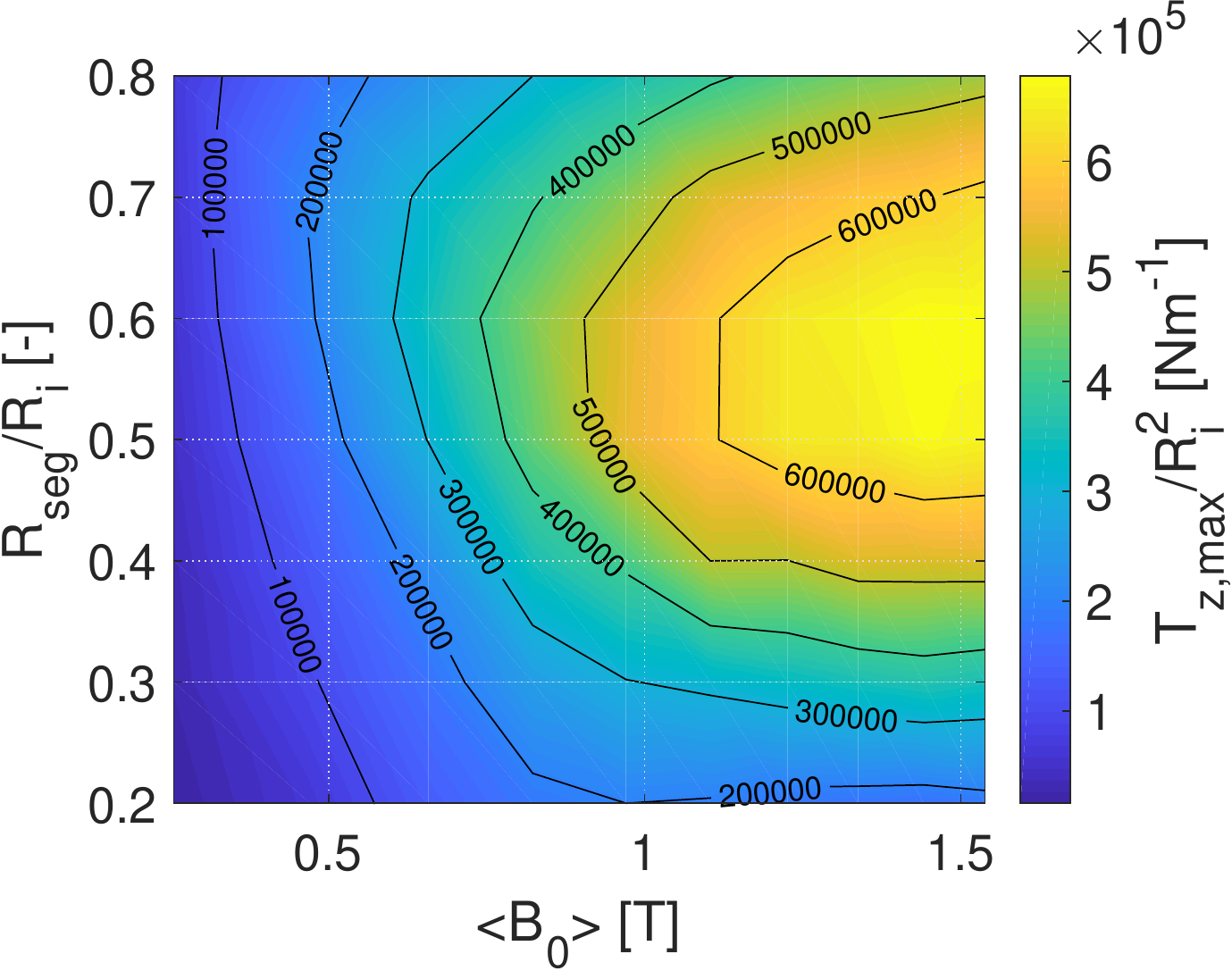}
\end{center}
\caption{b) the normalized dimensionalized torque profile as function of rotation angle for all geometries and b) the maximum dimensionalized torque per meter as function of the relative radius of the magnet and the switch.}\label{Fig_Torque}
\end{figure*}

\section{Torque}
Turning the inserts will require a torque, $T$. This has been computed for all geometries considered.  However, the computed torque depends on the dimensions of the geometry and must be dimensionalized for the results to be generally applicable. Here we have dimensionalized the torque by the inner radius of the Halbach cylinder. Thus when this parameter has been decided, the torque can be computed from the shown results by multiplying the shown value by this factor squared. Note that the computed torque is the torque per unit meter, as we are considering a two dimensional geometry. The normalized dimensionalized torque profile as function of angle for all geometries is shown in Fig \ref{Fig_Torque}a, while the maximum dimensionalized torque for all geometries is shown in Fig. \ref{Fig_Torque}b.

All torque profiles are relatively similar as can be seen in Fig. \ref{Fig_Torque}a, and in all cases the equilibrium positions, characterized by $T=0$, are located at the angles $\theta=[0^{\circ},90^{\circ},180^{\circ},270^{\circ}]$, corresponding to the low-field and high-field states. The stable equilibrium angles are the low-field states, when the reluctance of the magnetic circuit is minimized by diverting the field lines through the iron inserts. If the inserts are rotated towards positive angles, when crossing a stable equilibrium the torque goes from positive to negative. Vice-versa, the high-field states correspond to unstable equilibrium angles where the torque goes from negative to positive.

The negative of the torque profile reflected around any equilibrium angle is identical to the original torque profile, as expected from the $x$ and $y$ symmetries of the iron inserts.  However, the angles of maximum torque are not the same for all cases, although always in the interval  $(0^{\circ},45^{\circ}]$ and the three corresponding symmetrical positions. As can be noticed, the torque profiles are not symmetric around the points of maximum torque. When $R_\n{seg}/R_\n{i}$ is small and $R_\n{o}/R_\n{i}$ is large the torque profiles tend to be skewed in such a way that the modulus of the slope of $T(\theta)$ is higher around the unstable equilibrium angles than it is around the stable ones, indicating an energy landscape with wide stability valleys around the low-field states.

\section{Coercivity}\label{sec:Coercivity}
\begin{figure}[!t]
  \centering
  \includegraphics[width=1\columnwidth]{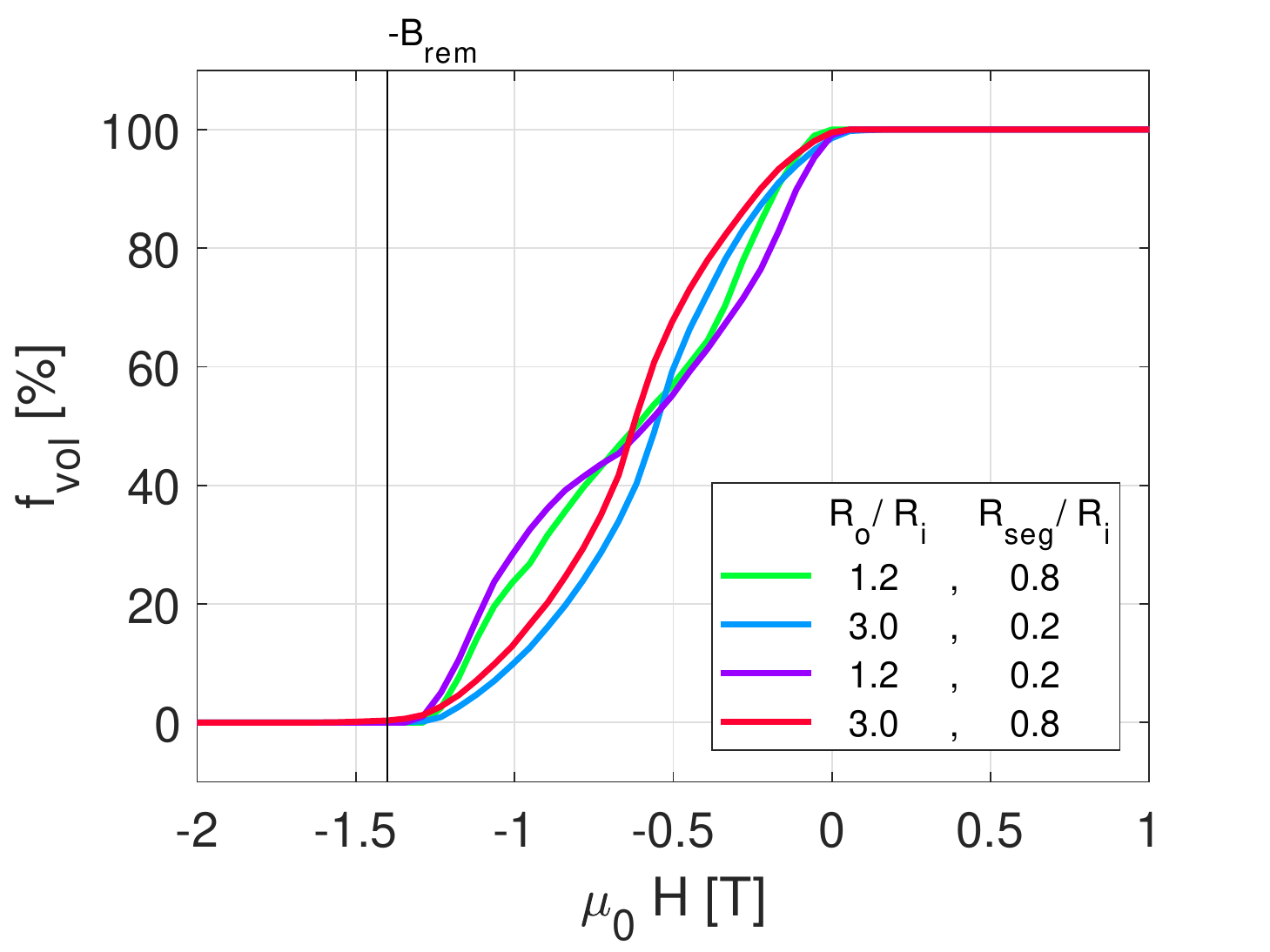}
  \caption{Volumetric distribution of operating points inside the permanent magnet material, showing that he demagnetizing field is safely above the values of coercivity of widely used grades of $\text{NdFeBo}$ magnets. The function $f_\n{vol}(H)$ is the fraction of magnet volume where $ H_{\parallel}<H$. }
  \label{Fig_SomeProfiles02}
\end{figure}

If the permanent magnet is subject to an intense demagnetizing field the operating point may move to the region of the $B$-$H$ curve where the linear model is no longer valid \cite{Insinga_2016a}. The presence of the iron inserts modifies the field distribution in any point of space, including inside the permanent magnet material. Therefore, it is necessary to verify our starting assumption that the permanent magnet  behaviour can be described by a linear constitutive relation. For this purpose we consider at any point of the magnet the direction of the remanence vector and we compute the component of the magnetic field along this direction. The parallel component of the field will be denoted by $H_{\parallel}$ and is defined as
\begin{eqnarray}
H_{\parallel}=(1/B_\n{rem})\big(H_{r} B_\n{rem,r}+H_{\phi}B_\n{rem,\phi} \big)
\end{eqnarray}
Non-linear demagnetization occurs when $H_{\parallel}$ is close to or less than the negative of the intrinsic coercivity, $H_c$, characterizing the permanent magnet material. We analysed the fraction $f_\n{vol}$ of permanent magnet volume where $\mu_0 H_{\parallel}$ is below a certain threshold value $H$, and calculated the results for values of $\mu_0 H$  spanning the interval $[-2 \text{T},1\text{T}]$. This gives the volumetric distribution of operating points inside the permanent magnet material for a given position of the iron inserts. Since rotating the inserts alters the distribution, we consider the maximum volume fraction $f_\n{vol}$ with respect to the angle $\theta$. Four examples are shown in Fig. \ref{Fig_SomeProfiles02}, corresponding to the four combinations of values with $R_\n{seg}/R_\n{i}=0.2$ or $0.8$, and $R_\n{o}/R_\n{i}=1.2$ or $3$. The volume fraction  $f_\n{vol}(H)$, here expressed in percentage, monotonically increases from $0\%$ to $100\%$, as the threshold field $H$ goes from $-2 \text{T}$ to $1 \text{T}$.

For standard grades of $\text{NdFeB}$ magnets the coercivity expressed in tesla is approximately as high as the remanence, $\mu_0 H_c \sim B_\n{rem}$, and for some grades it can be even higher than twice the remanence (UH and EH grades).  The fraction of permanent magnet volume where $\mu_0 H_{\parallel}<-B_\n{rem}$, denoted by $f_\n{vol}(-B_\n{rem})$, is thus a good indicator of the occurrence of non-linear demagnetization effects: if $f_\n{vol}(-B_\n{rem})$ remains very small, then the behaviour of the magnetic system is well described by the assumed linear relation. We verified that for all cases $f_\n{vol}(-B_\n{rem})$ is below $0.4 \%$, i.e. in the very worse case only 0.4\% of the Halbach magnet is demagnetized, if $\mu_0 H_c = B_\n{rem}$. A magnetic system realized with the most commonly used $\text{NdFeBo}$ grades would thus produce the field predicted by our calculations.

\begin{figure}[!t]
  \centering
  \includegraphics[width=1\columnwidth]{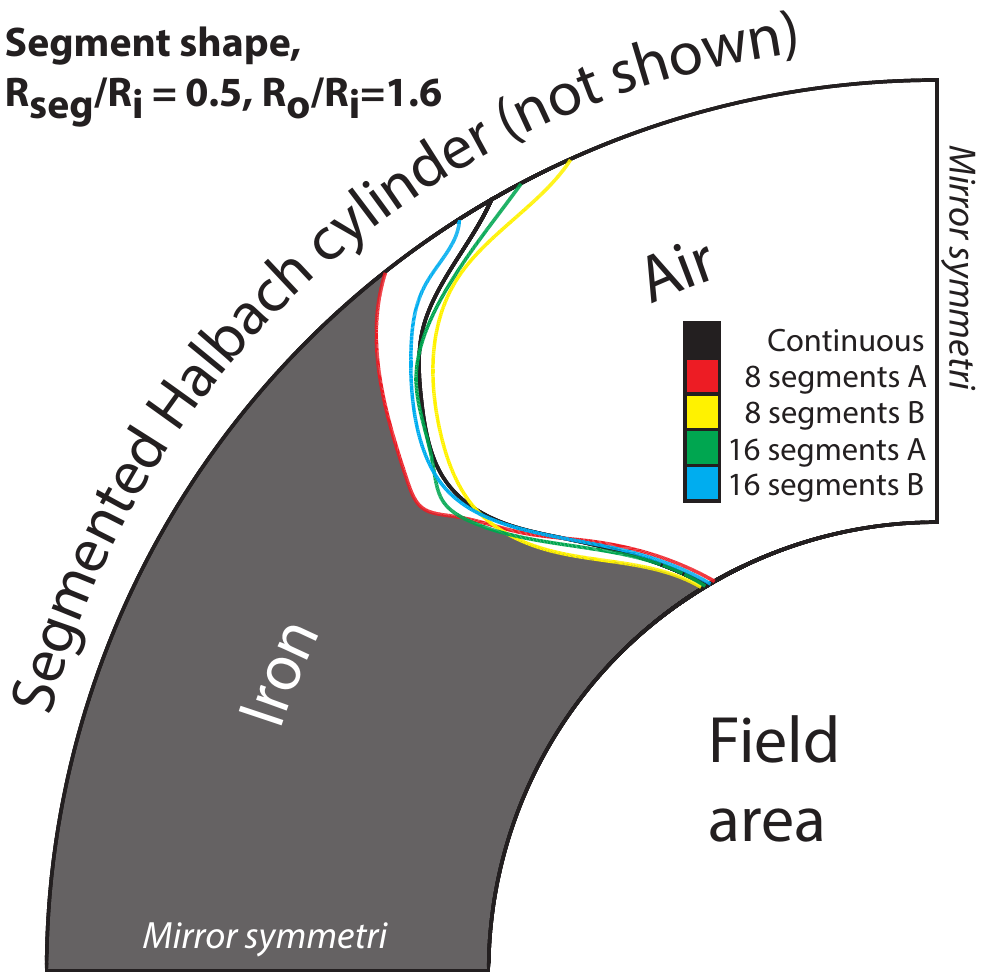}
  \caption{The shape of a quarter of the topology optimized segment as function of $R_\n{o}/R_\n{i}$ for $R_\n{seg}/R_\n{i}=0.6$, for four different segmentation of the Halbach cylinder field-source. There is mirror symmetry along the axes.}
  \label{Fig_Illus_switch_insert_segmented}
\end{figure}

\section{Segmented Halbach cylinder}\label{sec:Segmented}
All results discussed until now are for the continuous Halbach cylinder. Here we consider the more specific and more application oriented scenario of a segmented Halbach cylinder. We consider an eight and sixteen segmented cylinder, both in two configurations. In the first configuration, termed A, the segmentation is such that the first two segments are split by a line that is parallel to the field inside the cylinder bore. In the second configuration, termed B, the segments are rotated so that the first segment has a direction of magnetization that is parallel to field in the bore.

We only consider a specific geometry, which is the geometry with $R_\n{seg}/R_\n{i}=0.05$ and $R_\n{o}/R_\n{i}=1.6$ with $B_\n{rem}=1.4$ T, i.e. $B=B_\n{rem}\n{ln}(R_\n{o}/R_\n{i})=0.66$ T. For this geometry the topology optimized shape of the iron inserts are computed for the four segmented systems. For the case of configuration A with eight segments, the difference in field, i.e. $\langle{}B\rangle{}_{high}-\langle{}B\rangle{}_{low}$ is 0.64113 T while it is 0.65395 T for configuration B. For the case of 16 segments for configuration A the difference in field is 0.69355 T while it is 0.69705 T for configuration B. For the continuous case the difference is 0.71199 T. Thus the segmented geometries produce a difference in field that is almost that of the continuous case. The shape of the iron segment is shown in Fig. \ref{Fig_Illus_switch_insert_segmented} for the four segmented cases considered. As can be seen, the segment shape is almost identical to the continuous case, except for the 8 segmented system in configuration A, where the starting point of the segment is located closer to where the split between the two magnets are located.

The results for this studied segmented case suggests that the results for the continuous case studied in this work can be applied to the segmented Halbach cylinder, especially if the Halbach cylinder has 16 or more segments.

\section{Using a perfect magnetic insulator}
The topology optimized system so far have contained simple materials. We can however obtain a much larger difference in flux density by placing a material with $\mu_r=0$ in the air gap of the insert. Such a material can be realized using a superconducting material, see e.g. \cite{Navau_2012}. Having zero magnetic permeability, this material acts as a perfect magnetic insulator, and repels the field lines forcing them to pass through the air gap where the field intensity is thus increased. The exact shape of the insert for this type of material have to be recomputed. This has here only been done for the case of $R_\n{seg}/R_\n{i}=0.5$. Of course, the comparison of the topology optimized structure now also has to be done with a system with regular pole pieces with iron and $\mu_r=0$ structures.

\begin{figure}[!t]
  \centering
  \includegraphics[width=1\columnwidth]{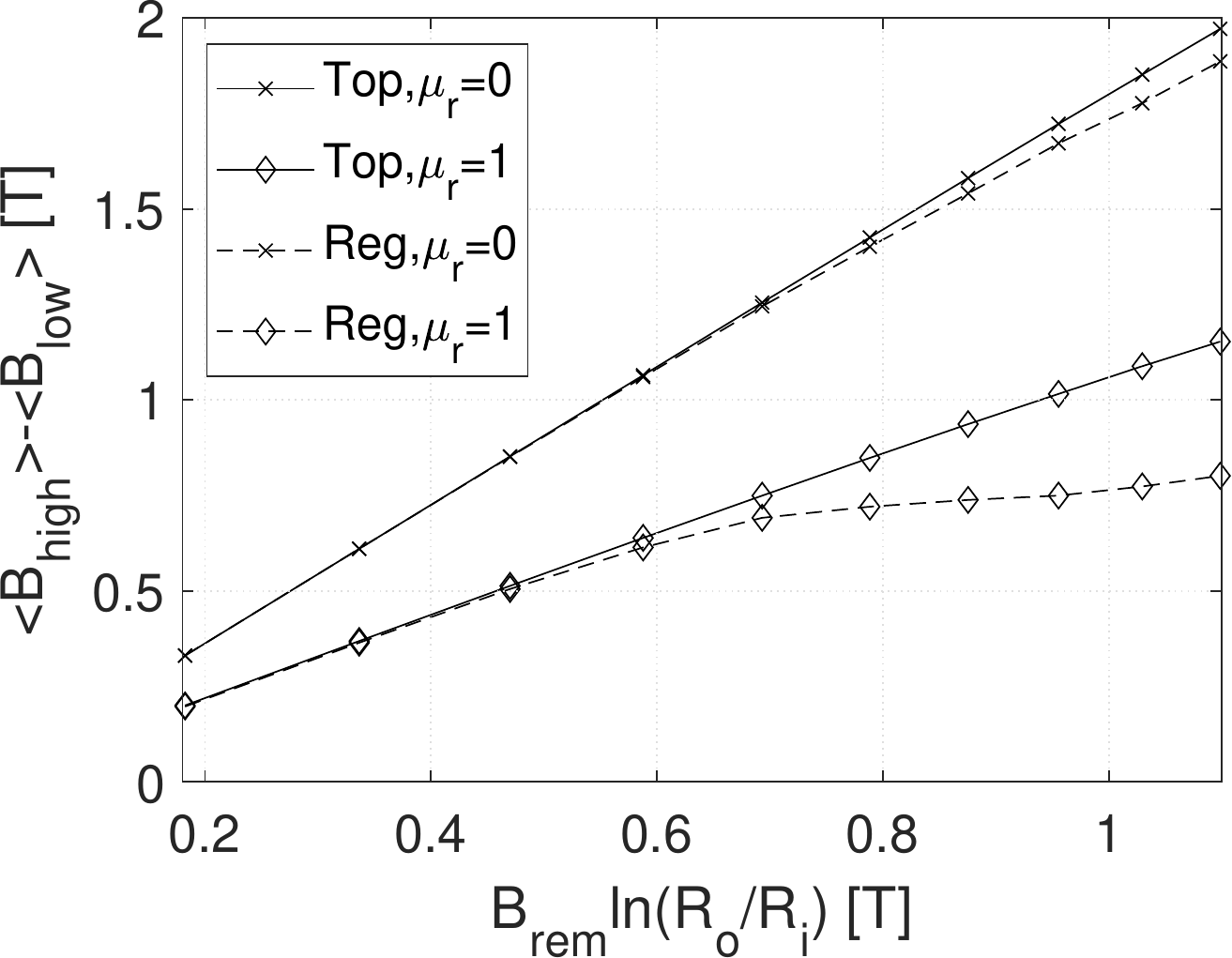}
  \caption{The difference in average high and low field of the topology optimized geometry and the regular pole pieces, for $\mu_r=0$ and 1, respectively.}
  \label{Fig_Mur0}
\end{figure}

The results for $\mu_r=0$ are shown in Fig. \ref{Fig_Mur0}. As can be seen, a magnetic structure with a much higher difference in field can be obtained if flux-repelling $\mu_r=0$ material is available. However, the regular pole pieces also perform much better in this case, leading to a much smaller difference between the topology optimized structures and the regular pole pieces, compared to the case for $\mu_r=1$, which are also shown in the figure.

\section{Conclusion}
We have investigated the design of a magnetic field source that can switch from a high field to a low field configuration by rotation by $90^\circ$ of a set of iron pieces within the air gap of a Halbach cylinder. Using topology optimization, the ideal shape of the iron inserts was determined as function of the geometrical parameters of the system. It was shown that the topology optimized structures have a difference in flux density between the high and low configurations that is on average 1.29 times higher than optimized regular pole pieces. The maximum increase is a factor of 2.08 times higher than regular pole pieces.

\section{Appendix}
\begin{figure}[!b]
  \centering
  \includegraphics[width=0.97\columnwidth]{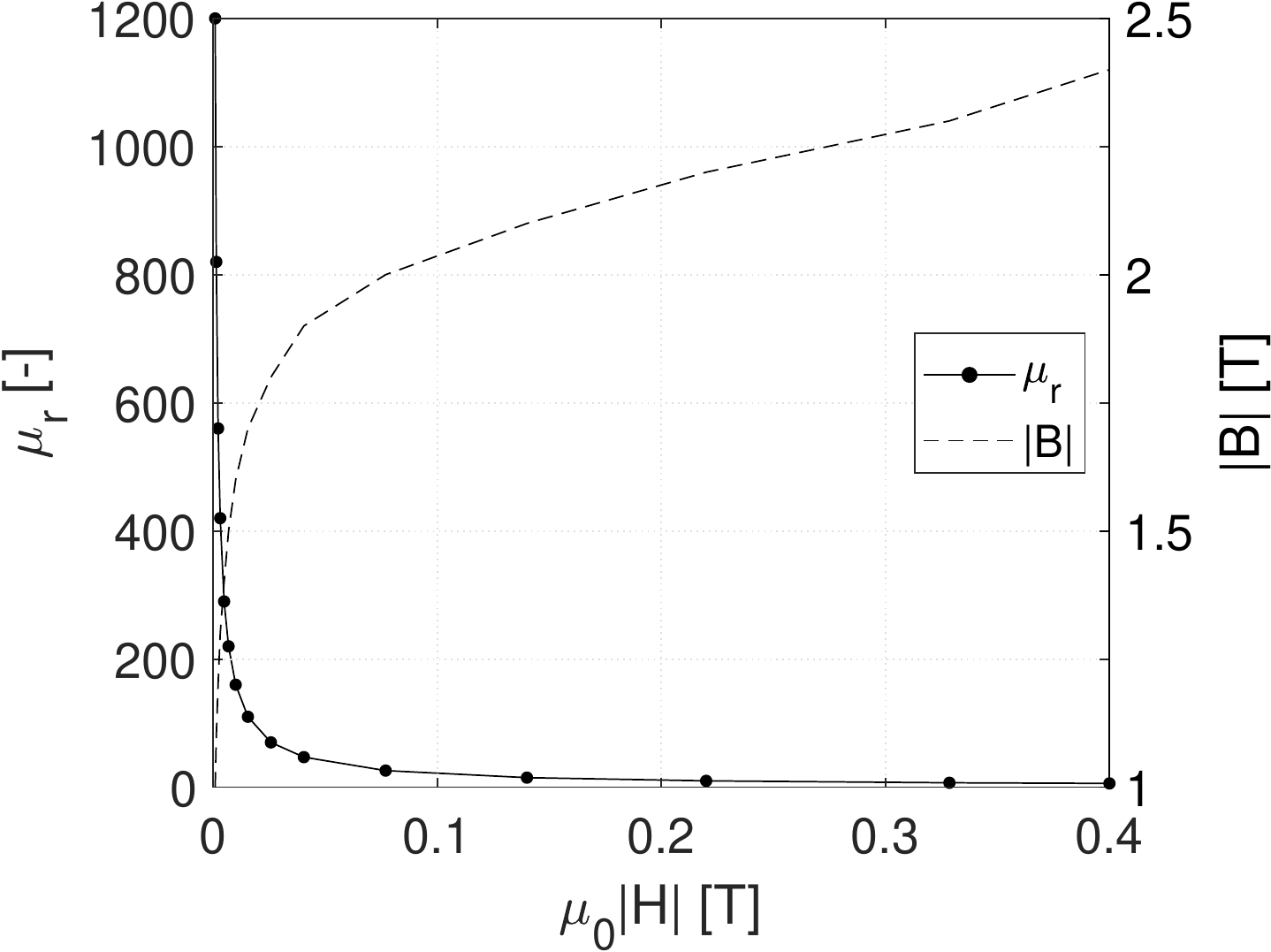}
  \caption{The relative permeability, $\mu_r$, and the corresponding norm of the magnetic flux density, $|B|$, as function of the norm of the magnetic field, $|H|$, for the iron as used throughout this work. Outside the plotted range of $|H|$ the values are assumed constant with the same value as the last data point. The data is taken from the Comsol material library.}
  \label{Fig_Iron_curve_mur}
\end{figure}

Shown in Fig. \ref{Fig_Iron_curve_mur} is the relative permeability data as function of magnetic field for iron that is used throughout this work.

\begin{figure}[!b]
  \centering
  \includegraphics[width=0.97\columnwidth]{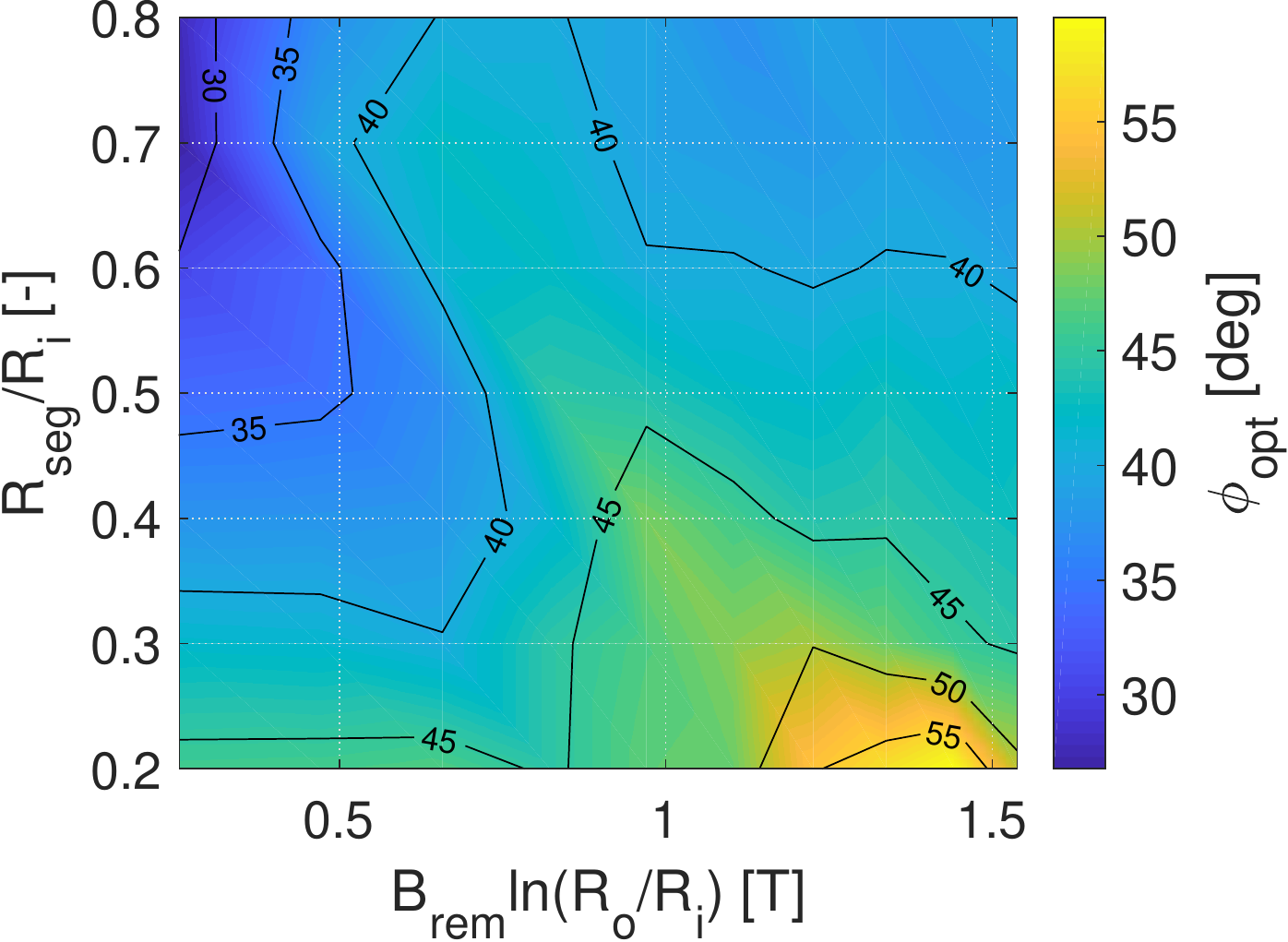}
  \caption{The angular extent of the iron segments for the regular geometry, the angle $\phi$ shown in Fig. \ref{Fig_Switch_Halbach_ill}, as function of the relative radius of the magnet and the switch.}
  \label{Fig_Angle_14_cut}
\end{figure}

The optimal angle as function of  the ratio between $R_\n{seg}$ and $R_\n{i}$ and the field in the central bore for the regular pole pieces is shown in Fig. \ref{Fig_Angle_14_cut}.

\clearpage

\bibliographystyle{elsarticle-harv}

\end{document}